\documentclass[pra,a4paper,oneside,twocolumn,superscriptaddress,showpacs]{revtex4-1}

\usepackage[english]{babel}
\usepackage[latin1]{inputenc}
\usepackage{dcolumn}
\usepackage{subfigure}
\usepackage{multirow}
\usepackage{amsmath,amsfonts,amssymb}
\usepackage{graphicx}  
\usepackage{latexsym}   
\usepackage{color, ulem}

\begin{document}


\title{Controlling optical memory effects in disordered media with coated metamaterials}

\author{\firstname{Tiago}  J. \surname{Arruda}}
\email{tiagojarruda@gmail.com}
\affiliation{Instituto de F\'isica de S\~ao Carlos,
Universidade de S\~ao Paulo, 13566-590 S\~ao Carlos-SP, Brazil}

\author{\firstname{Alexandre} S. \surname{Martinez}}
\affiliation{Faculdade de Filosofia,~Ci\^encias e Letras de Ribeir\~ao
Preto, Universidade de S\~ao Paulo, 14040-901 Ribeir\~ao
Preto-SP, Brazil}
\affiliation{National Institute of Science and Technology in
Complex Systems, 22290-180 Rio de Janeiro-RJ, Brazil}

\author{\firstname{Felipe}  A. \surname{Pinheiro}}
\affiliation{Instituto de F\'{i}sica, Universidade Federal do Rio de Janeiro, 21941-972 Rio de Janeiro-RJ, Brazil}


\begin{abstract}
Most applications of memory effects in disordered optical media, such as the tilt-tilt and shift-shift spatial correlations, have focused on imaging through and inside biological tissues. Here we put
forward a metamaterial platform not only to enhance but also to tune memory effects in random media. Specifically, we investigate the shift-shift and tilt-tilt spatial correlations in metamaterials composed of coated spheres and cylinders by means of the radiative transfer equation.
Based on the single-scattering phase function, we calculate the translation correlations in anisotropically scattering media with spherical or cylindrical geometries and find a simple relation between them.
We show that the Fokker-Planck model can be used with the small-angle approximation to obtain the shift-tilt memory effect with ballistic light contribution.
By considering a two-dimensional scattering system, composed of thick dielectric cylinders coated with subwavelength layers of thermally tunable magneto-optical semiconductors, we suggest the possibility of tailoring and controlling the shift-shift and tilt-tilt memory effects in light scattering.
In particular, we show that the generalized memory effect can be enhanced by increasing the temperature of the system, and it can be decreased by applying an external magnetic field.
Altogether our findings unveil the potential applications that metamaterial systems may have to control externally memory effects in disordered media.
\end{abstract}



\maketitle


\section{Introduction}

The usual memory effect in light scattering refers to the short-range angular correlation between a rotated incident beam over small angles and the resulting speckle pattern~\cite{Feng_PhysRevLett61_1988,Stone_PhysRevLett61_1988,Freund_PhysA168_1990,Genack_PhysRevE49_1994}.
 This memory effect shows up by tilting the electromagnetic field incident upon a diffuser, leading to a corresponding tilting of the scattered field by the same amount, which is limited by a certain angular-correlation range~\cite{Feng_PhysRevLett61_1988,Snieder_EPJ226_2017}.
These spatial correlations find applications in areas where knowledge of the intensity transmission matrices of diffusers is crucial, ranging from optical communications and wavefront shaping~\cite{Freund_PhysA168_1990,Fink_NatPhot6_2012} to adaptive optics~\cite{Bifano_ApplOpt54_2015}, and specially in biomedical imaging of structures within soft tissues~\cite{Mosk_Nature491_2012,Silberberg_NatPhot6_2012,Gigan_NatPhot8_2014,Psaltis_OptExp22_2014,Psaltis_OptExp23_2015}.

Recently, a new class of correlations termed the {\it shift-shift} memory effect has been shown to be the Fourier complement of the traditional {\it tilt-tilt} memory effect in light scattering~\cite{Judkewitz_NatPhys11_2015}.
These shift-shift correlations, which imply a translation-invariant speckle pattern, do not impose thin scattering layers at a distance from the target, and can be directly determined during plane illumination from the resulting spatial speckle autocorrelation function.
Recently it has been shown that the tilt-tilt and shift-shift memory effects actually are special cases of a more general class of {\it shift-tilt} spatial correlations, as a result of the combination of short-range angular and translation correlations in scattering media exhibiting forward scattering~\cite{Vellekoop_Optica4_2017}.
This {\it generalized memory effect} allows for shifting and tilting of a focal spot formed deep within anisotropically scattering media in an optimal joint shift-tilt scheme, which finds applications in recording of clear images from within biological tissues~\cite{Mosk_Nature491_2012,Silberberg_NatPhot6_2012,Gigan_NatPhot8_2014,Vellekoop_Optica4_2017}.

Most applications of optical memory effects in disordered scattering media have been focused on imaging through and deep within complex dielectric media, especially in biomedical imaging~\cite{Judkewitz_NatPhys11_2015,Vellekoop_Optica4_2017}.
Conversely, little attention has been devoted to optical memory effects in artificial scattering media, in which these spatial correlations could be potentially enhanced and controlled.
With this motivation in mind, in the present paper we investigate the shift-tilt memory effect in metamaterials composed of core-shell scatterers with spherical or cylindrical geometries.
Core-shell particles are known to exhibit interesting single-scattering responses such as Fano resonances~\cite{Arruda_PhysRevA87_2013,Arruda_PhysRevA92_2015,Zayats_OptExp21_2013,Gao_OptExp21_2013,Limonov_SciRep5_2015,Arruda_PhysRevA96_2017}, plasmonic cloaking~\cite{Alu2008,Monticone_PhysRevLett110_2013,Kort-Kamp_PhysRevLett111_2013}, enhanced stored energy~\cite{Arruda_JOpt14_2012,Arruda_JOpSocAmA31_2014,Arruda_JOSA32_2015}, superscattering~\cite{Kivshar_OptLett38_2013,Gao_JPhysC118_2014}, and near-zero-forward or zero-backward scattering of light~\cite{Kivshar_ACSNano6_2012,Wang_SciRep5_2015}.
In particular, we have recently shown that the scattering anisotropy of a collection of magneto-optical core-shell microcylinders can be externally tuned by applying a moderate magnetic field in the terahertz frequency range~\cite{tiago-pra2016}.
Here, by using a system composed of thick dielectric cylinders coated with subwavelength magneto-optical and thermally tunable semiconductors, we demonstrate one can tune and tailor the shift-tilt memory effect by externally applying a magnetic field or by varying the temperature.

This paper is organized as follows.
In Sec.~\ref{sec2:theory}, we provide an overview of the shift-shift correlation function calculated within the small-angle approximation (SAA) for large spheres and derive a similar expression to be used for cylindrical scatterers.
In Sec.~\ref{sec21:generalized}, we briefly discuss the generalized memory effect derived from the Fokker-Planck light propagation model and propose an effective distance shift to be used in the SAA.
 A comparison of spatial correlation functions in dielectric anisotropically scattering media containing spherical versus cylindrical scatterers is provided in Sec.~\ref{sec3:dielectric-cylinders}.
Finally, in Sec.~\ref{sec4:magneto-optical-cylinders}, we show the possibility of tuning the anisotropic memory effect associated with coated scatterers in the terahertz range by varying the temperature or by applying an external magnetic field.
We summarize our main results in Sec.~\ref{sec5:conclusion}.

\section{Anisotropic memory effect}

At time $t$, for two speckle patterns associated with the complex electric field $\mathbf{E}(\mathbf{r},t)$ at positions $\mathbf{r}_1$ and $\mathbf{r}_2$, the optical memory effect is quantified by the spatial coherence function (also referred to as the mutual intensity):
\begin{align}
\Gamma(\mathbf{r}_1,\mathbf{r}_2,t)=\langle\mathbf{E}(\mathbf{r}_1,t)\cdot\mathbf{E}^*(\mathbf{r}_2,t)\rangle,\label{Gamma-0}
\end{align}
where $\langle\ldots\rangle$ implies an ensemble average over all possible realizations of $\mathbf{E}(\mathbf{r},t)$~\cite{Freund_PhysA168_1990,Raymer_PhysRevA62_2000}.
The details of how to measure and calculate the shift-shift memory effect in light scattering by using transmission matrices, and its shift-shift and/or tilt-tilt generalization using a paraxial model of light propagation, are discussed in Refs.~\cite{Judkewitz_NatPhys11_2015,Vellekoop_Optica4_2017}.
Here, we are interested in the application of these spatial correlations to metamaterials consisting of three- and two-dimensional disordered media.
Our aim is to exploit the single-particle scattering solution for spheres and cylinders to achieve anisotropically scattering media with tunable shift-tilt memory effect.

\subsection{Small-angle approximation: Shift-shift correlations for monodispersed spherical or cylindrical Mie scatterers}
\label{sec2:theory}

Let us consider light scattering in the quasiballistic regime, in which multiple scattering quantities can be derived by the single-particle scattering phase function~\cite{ishimaru,Arruda_JOSAA34_2017}.
As discussed in Ref.~\cite{Judkewitz_NatPhys11_2015}, the translation or shift-shift correlation function in highly anisotropic scattering media can be accurately calculated in the framework of the Lorenz-Mie theory~\cite{ishimaru,bohren}.
Indeed, for near-forward-scattering systems, the small-angle approximation of the radiative transfer (RT) equation provides accurate shift-shift spatial correlation functions in the quasiballistic regime~\cite{Judkewitz_NatPhys11_2015}, so that one can apply single-scattering phase functions based on the Lorenz-Mie theory to describe ensemble-averaged scattering properties.

Since Eq.~(\ref{Gamma-0}) for $\mathbf{r}_1=\mathbf{r}_2$ may produce several definitions of single-scattering phase functions, let us define the phase function $p(\theta)$ as the probability distribution function describing the scattered light angular distribution, so that  $\int_{4\pi}{\rm d}\Omega p_{\rm s}(\theta)=1$ for spherical particles and $\int_{0}^{2\pi}{\rm d}\theta p_{\rm c}(\theta)=1$ for cylindrical particles~\cite{ishimaru,bohren}.
The spatial correlation between two scattered fields, $\mathbf{E}_{\rm sca}(\mathbf{r})$ and $\mathbf{E}_{\rm sca}(\mathbf{r}-\boldsymbol{\Delta}\mathbf{r})$, of a system of uncorrelated arbitrary particles can be calculate from the ensemble average~\cite{ishimaru}
\begin{align}
\langle \mathbf{E}_{\rm sca}(\mathbf{r})\cdot \mathbf{E}_{\rm sca}^*(\mathbf{r}-\boldsymbol{\Delta}\mathbf{r})\rangle\propto I_0\int_V {\rm d}V \rho p(\theta) \exp(\imath\mathbf{k}\cdot\boldsymbol{\Delta}\mathbf{r}),\label{spatial}
\end{align}
where $I_0$ is the intensity of the incident wave and $\rho$ is the density of scatterers.
Here, $\Delta r=|\boldsymbol{\Delta}\mathbf{r}|$ is limited within the order of the correlation distance and $|\mathbf{r}|\gg \Delta r$, so that one can define $p(\theta)$ within the integral~\cite{ishimaru}.
For spherical waves $(|\mathbf{E}_{\rm sca}^{\rm s}|^2\propto1/r^2)$ in spherical coordinates $(r,\theta,\varphi)$ with $\boldsymbol{\Delta}\mathbf{r}=\Delta r\hat{\mathbf{x}}$, one has ${\rm d}V \exp(\imath\mathbf{k}\cdot\boldsymbol{\Delta}\mathbf{r})=r^2{\rm d}r{\rm d}\Omega\exp(\imath\alpha \sin\theta\cos\varphi)$, where $\alpha=k\Delta r$ and $k=|\mathbf{k}|$.
Considering $|\theta|\ll1$ (i.e., $\sin\theta\approx\theta$) and recalling that $\int_0^{2\pi}{\rm d}\varphi\exp(\imath\alpha\theta\cos\varphi)=2\pi J_0(\alpha\theta)$, where $J_0$ is the zeroth-order Bessel function, Eq.~(\ref{spatial}) retrieves the Fourier-Bessel transform of a forward-extended $p_{\rm s}(\theta)$~\cite{ishimaru}.
Conversely, for cylindrical waves $(|\mathbf{E}_{\rm sca}^{\rm c}|^2\propto1/r)$ in cylindrical coordinates $(r,\varphi,z)$ with $\boldsymbol{\Delta}\mathbf{r}=\Delta r\hat{\mathbf{y}}$, one has ${\rm d}V \exp(\imath\mathbf{k}\cdot\boldsymbol{\Delta}\mathbf{r})=r{\rm d}r{\rm d}\theta{\rm d}z\exp(\imath \alpha \sin\theta)$, where $\theta=\varphi-\pi$.
Imposing $|\theta|\ll1$, it follows that we must consider a Fourier transform of $p_{\rm c}(\theta)$ in polar variables instead of the Fourier-Bessel transform.
Both operations, however, are equivalent to a two-dimensional Fourier transform reduced by symmetry in the $\varphi$ direction.
Explicitly,  with $p(\theta)\approx 0$ for $|\theta|\geq\pi/2$, we can define for small scattering angles the Fourier transforms
\begin{align}
\widetilde{p}_{\rm s}(\alpha) &\equiv 2\pi\int_0^{\infty} {\rm d}\theta p_{\rm s}(\theta)J_0(\alpha\theta)\theta,\label{trans-sphere}\\
\widetilde{p}_{\rm c}(\alpha) &\equiv 2\int_0^{\infty} {\rm d}\theta p_{\rm c}(\theta) \cos(\alpha\theta).\label{trans-cylinder}
\end{align}

Equation~(\ref{trans-sphere}) is valid for media whose scattering phase functions are forward-extended, so that the small-angle approximation of the RT equation can be applied~\cite{kokhanovsky,kokhanovsky_book}.
Here, we propose the use of Eq.~(\ref{trans-cylinder}) for cylindrical scattered waves, i.e., disordered media composed of parallel cylindrical scatterers normally irradiated with plane waves.
Since the RT equation depends not on the shape of scatterers but on the ensemble averages of intensities, Eq.~(\ref{trans-cylinder}) is the only critical modification to enter our calculations.
As a result, for both  spherical and cylindrical monodispersed scatterers, the transverse coherence function (mutual intensity) for a shift distance $\Delta r$ in a target plane perpendicular to the incidence direction $\mathbf{k}_0$ is~\cite{ishimaru,kokhanovsky,Judkewitz_NatPhys11_2015}
\begin{align}
\Gamma(\alpha)=I_0\exp\left\{-\tau\left[1-\varpi_0\widetilde{p}(\alpha)\right]\right\},\label{correlation0}
\end{align}
where $\tau=L/\ell_{\rm ext}$ is the optical thickness, with $\ell_{\rm ext}=1/\rho \sigma_{\rm ext}$, $\alpha=k\Delta r$ and $\widetilde{p}(\alpha)$ can be either Eqs.~(\ref{trans-sphere}) or (\ref{trans-cylinder}) for spherical~\cite{ishimaru,gaskill} or cylindrical scatterers, respectively, with $\varpi_0=\sigma_{\rm sca}/\sigma_{\rm ext}$ being the corresponding single scattering albedo.
The quantity $\Gamma(\alpha)$ can be identified as the two-dimensional Fourier transform of the point-spread function~\cite{swanson}, which is associated with loss of resolution in images taken through scattering media~\cite{Judkewitz_NatPhys11_2015}.
As $\Gamma(\alpha)$ is maximal for $\Delta r=0$, the shift-shift correlation function can be defined as $\mathcal{C}^{\rm RT}(\alpha)\equiv\Gamma(\alpha)/\Gamma(0)$, which leads to
\begin{align}
\mathcal{C}^{\rm RT}(\alpha)=\exp\left\{-\tau\varpi_0\left[1-\widetilde{p}(\alpha)\right]\right\}.\label{correlation}
\end{align}
Equation~(\ref{correlation}) is calculated in Ref.~\cite{Judkewitz_NatPhys11_2015} for large dielectric spheres ($kR\gg1$) and a low density of scatterers, showing very good agreement with experimental and simulation data.
The only approximation made is related to the scattering anisotropy: $\langle\cos\theta\rangle\approx 1$.
As $\langle\cos\theta\rangle \to 1$, the normalization of the phase function implies that $\widetilde{p}(0)\to1$.
The approximate expressions for the phase functions of large spheres and thick cylinders and their corresponding Fourier transforms are provided in the Appendix.

As $\widetilde{p}(\alpha)\approx 0$ which occurs at $\alpha>2kR$ for spherical or cylindrical scatterers (see the Appendix), Eq.~(\ref{correlation}) reduces to a constant background $\exp(-\varpi_0\tau)$ in the correlations, which depends on the shape of scatterers and decays exponentially as the scattered light becomes stronger.
This constant background appears for optical thickness $\tau<1$ and is due to speckles that are not yet fully developed in the quasiballistic regime~\cite{Judkewitz_NatPhys11_2015}.
Indeed, for both spherical and cylindrical waves, $\Gamma(\alpha)$ depends only on $\tau$ and $\varpi_0$ when $\alpha>2kR$ and $kR\gg1$.
Using the extinction paradox, it follows that $Q_{\rm ext}\equiv\sigma_{\rm ext}/\sigma_{\rm g}\to2$ as $kR\to\infty$, where $\sigma_{\rm g}$ is the geometric cross section and $Q_{\rm ext}$ is the extinction efficiency.
Recalling the definition of optical thickness, we have $\tau_{\rm s}=L/\ell_{\rm ext}^{\rm s}=(3/4) f Q_{\rm ext}^{\rm s}L/R$ for spheres and $\tau_{\rm c}=(2/\pi) f Q_{\rm ext}^{ c}L/R$ for cylinders, where $f$ is the packing fraction of scatterers.
The application of the extinction paradox and the condition $\alpha>2kR$ for weakly dissipative scatterers ($\varpi_0\approx 1)$ allow us to write a general relation between the transverse coherence function for monodispersed large spheres and that for thick cylinders,
\begin{align}
\Gamma_{\rm s}(\alpha)\approx \left[\Gamma_{\rm c}(\alpha)\right]^{3\pi/8},\label{relation}
\end{align}
since $\tau_{\rm s}/\tau_{\rm c}\approx 3\pi/8$.
The main consequence of Eq.~(\ref{relation}) is that the constant background $\Gamma(\alpha>2kR)$ associated with cylindrical scatterers is slightly larger than that for spheres, provided the optical parameters and $k$, $R$, $L$ and $\rho$ are the same.

Equation~(\ref{relation}) was derived by invoking geometric arguments since $4R/3$ and $\pi R/2$ are obtained from the ratio between the scatterer volume $v$ and its geometric cross section $\sigma_{\rm g}$~\cite{Arruda_JOSAA34_2017}.
 Hence we conclude that monodispersed scatterers with a ratio $kv/\sigma_{\rm g}$ higher than $kR\pi/2$ should show greater values of transverse coherence than monodispersed cylinders with equivalent parameters.
For monodispersed large scatterers of arbitrary shapes, we may find a constant background derived from the transverse coherence function that satisfies
\begin{align}
\left[\frac{\Gamma(k\Delta r)}{I_0}\right]^{kv/\sigma_{\rm g}}\approx \exp(-2fkL) \label{C-approx}
\end{align}
when $Q_{\rm ext}=2$, $\Delta r> 2 R_{\rm eff}$ and $\langle\cos\theta\rangle\approx1$, where $R_{\rm eff}$ is the effective radius of the scatterers.
Equation~(\ref{C-approx}) is an original result of our study and shows explicitly that the constant background from the transverse coherence function can be obtained by geometric parameters alone.

\subsection{Fokker-Planck model of light propagation: Generalized optical memory effect for lossless dielectric scatterers}
\label{sec21:generalized}

Recently, a generalization of the optical memory effect that takes into account both shift-shift and tilt-tilt anisotropic memory effects was reported~\cite{Vellekoop_Optica4_2017}.
This generalization allows one to perform a shift-tilt $(\boldsymbol{\Delta}\mathbf{r}_0,\boldsymbol{\Delta}\mathbf{k}_0)$ of the incident field on the input plane in order to attain a specific shift-tilt $(\boldsymbol{\Delta}\mathbf{r},\boldsymbol{\Delta}\mathbf{k})$ of the speckle on the target plane.
To describe this shift-tilt generalized correlation function, a paraxial light propagation model based on the Fokker-Planck equation was developed and verified experimentally~\cite{Vellekoop_Optica4_2017}.
The FP model is based on a simplification of the Wigner-function transport equation~\cite{Raymer_PhysRevA62_2000}, which neglects the contribution of the ballistic (unscattered) light to the two-point spatial coherence function.
The approximate solution does not impose any restriction on the shape of scatterers and is valid for lossless near-forward-scattering materials with $L>\ell_{\rm tr}$, where $\ell_{\rm tr}$ is the transport mean free path.
Explicitly, the spatial correlation function is a functional of two vector quantities and is given by~\cite{Vellekoop_Optica4_2017}
\begin{align}
&\mathcal{C}^{\rm FP}(\boldsymbol{\Delta}\mathbf{r},\boldsymbol{\Delta}\mathbf{k})\nonumber\\
&=\exp\left[-\frac{L^3k^2}{2\ell_{\rm tr}}\left( \frac{|\boldsymbol{\Delta}\mathbf{k}|^2}{3k^2}-\frac{\boldsymbol{\Delta}\mathbf{k}\cdot\boldsymbol{\Delta}\mathbf{r}}{kL}+\frac{|\boldsymbol{\Delta}\mathbf{r}|^2}{L^2} \right)\right],\label{correlation-FP}
\end{align}
where $\ell_{\rm tr}=1/[\rho\sigma_{\rm sca}(1-\langle\cos\theta\rangle)]$, with $\langle\cos\theta\rangle$ being the single-scattering asymmetry parameter.
One can verify that $\boldsymbol{\Delta}\mathbf{r}=\boldsymbol{\Delta}\mathbf{r}_0 + L\boldsymbol{\Delta}\mathbf{k}_0/k$ and $\boldsymbol{\Delta}\mathbf{k}=\boldsymbol{\Delta}\mathbf{k}_0$, which are expected results from the ballistic propagation of light through a very dilute transparent medium of thickness $L$.

Interestingly, the additional offset $L\boldsymbol{\Delta}\mathbf{k}/k$ in the target intensity distribution suggests that the SAA of the radiative transfer equation, given in Eq.~(\ref{correlation}), could be readily applied to this general case of tilt-tilt and/or shift-shift correlations by using an ``effective distance shift''.
A  possible substitution into Eq.~(\ref{correlation}) would be $\alpha\to{k}|\boldsymbol{\Delta}\mathbf{r}_0 + L\boldsymbol{\Delta}\mathbf{k}_0/k|$ for the case in which $\boldsymbol{\Delta}\mathbf{k}$ is in a plane perpendicular to the incident direction.
However, as it can be readily verified, $\boldsymbol{\Delta}\mathbf{r}_0 = - L\boldsymbol{\Delta}\mathbf{k}_0/k$ would imply $\alpha=0$, which is not a correct result since ``shifting'' and ``tilting'' the incident wave with respect to the media correspond to different symmetry operations.
This suggests that the appropriate effective function should have the form $\alpha^2={k}^2|\boldsymbol{\Delta}\mathbf{r}_0 + L\boldsymbol{\Delta}\mathbf{k}_0/k|^2 +\eta^2$, where $\eta=\eta(\boldsymbol{\Delta}\mathbf{r}_0,\boldsymbol{\Delta}\mathbf{k}_0)\not=0$ for $\boldsymbol{\Delta}\mathbf{r}_0 = - L\boldsymbol{\Delta}\mathbf{k}_0/k$ plays the role of a nonlinear perturbation.
This heuristic approach to finding the approximate $\alpha$ to be used in the SAA is convenient, since the rigorous solution of the RT equation taking into account both ``tilting'' and ``shifting'' would lead to a very complicated analytical form.
In addition, for dispersive metamaterials we cannot apply Eq.~(\ref{correlation-FP}), which provides accurate results only for lossless dielectric scatterers.

To find the correct effective distance shift $\alpha$ to be used in the SAA solution, we compare Eq.~(\ref{correlation-FP}) to  Eq.~(\ref{correlation}).
By inspection, it seems convenient to rewrite the paraxial correlation function $\mathcal{C}^{\rm FP}$ as a function of the dimensionless variable $\alpha=\alpha(\boldsymbol{\Delta}\mathbf{r},\boldsymbol{\Delta}\mathbf{k})$.
Using an appropriate notation, we rewrite Eq.~(\ref{correlation-FP}) as
\begin{align}
\mathcal{C}^{\rm FP}(\alpha)=\exp\left(-\frac{L\alpha^2}{2\ell_{\rm tr}}\right), \label{CFP-alpha}
\end{align}
where we have defined
\begin{align}
\alpha\equiv kL\sqrt{\frac{|\boldsymbol{\Delta}\mathbf{k}|^2}{3k^2}-\frac{\boldsymbol{\Delta}\mathbf{k}\cdot\boldsymbol{\Delta}\mathbf{r}}{kL}+\frac{|\boldsymbol{\Delta}\mathbf{r}|^2}{L^2}}.\label{alpha-effective}
\end{align}
Equation~(\ref{alpha-effective}) is the correct approximate effective distance shift to be used in our heuristic approach to the RT equation in the SAA.
To demonstrate this, let us consider Eq.~(\ref{trans-sphere}) for $L>\ell_{\rm tr}$.
For $|\alpha\theta|\ll1$, one has the approximation $J_0(\alpha\theta)\approx 1-(\alpha\theta)^2/4$, which leads to  $\widetilde{p}_{\rm s}(\alpha)\approx 2\pi\int_0^{\pi}{\rm d}\theta p_{\rm s}(\theta)[1-(\alpha\theta)^2/4]\theta$.
Using $\cos\theta\approx 1-\theta^2/2$ and recalling the definition of the scattering asymmetry parameter $\langle\cos\theta\rangle$, we finally obtain
\begin{align}
\widetilde{p}_{\rm s}(\alpha)\approx 1 - \left(1-\langle\cos\theta\rangle\right)\frac{\alpha^2}{2}. \label{phase-approx}
\end{align}
Substituting Eq.~(\ref{phase-approx}) into Eq.~(\ref{correlation}) and considering $\varpi_0\approx1$ (and hence $\ell_{\rm ext}\approx\ell_{\rm sca}$), we retrieve Eq.~(\ref{CFP-alpha}), but now in the context of the radiative transfer equation in the SAA.
The same result can be obtained from Eq.~(\ref{trans-cylinder}) by using an appropriate upper limit $\theta_{\rm max}$ for the angular integral.

As expected, $\mathcal{C}^{\rm FP}$ is an approximation for $\mathcal{C}^{\rm RT}$ when $\exp(-\tau)\approx 0$, since the FP model does not include ballistic light~\cite{Vellekoop_Optica4_2017}.
Note that a similar expression for $\alpha$ could be obtained in the SAA using simple geometric arguments.
If we fix $\boldsymbol{\Delta}\mathbf{r}=\Delta r\hat{\mathbf{x}}$ and the incident direction along the $z$ axis, while keeping $\boldsymbol{\Delta}\mathbf{k}$ as a free parameter, the approximate distance shift in the $xy$ plane would be $|\Delta r\hat{\mathbf{x}}-L\Delta k(\hat{\mathbf{x}}+\hat{\mathbf{y}})/k\sqrt{3})|$.

Henceforth, we focus our attention on $\mathcal{C}^{\rm RT}(\alpha)$, Eq.~(\ref{correlation}), evaluated at the effective distance shift given by Eq.~(\ref{alpha-effective}).
Of course, for $\boldsymbol{\Delta}\mathbf{k}=\mathbf{0}$, Eq.~(\ref{alpha-effective}) retrieves the shift-shift memory effect with $\alpha=k\Delta r$.
More importantly, Eq.~(\ref{alpha-effective}) leads to the same optimal isoplanatic patch verified experimentally~\cite{Vellekoop_Optica4_2017}, which is a small area where the focus scan range is limited in adaptive optics applications~\cite{Bifano_ApplOpt54_2015}.
The optimal isoplanatic patch for a desired distance shift $\boldsymbol{\Delta}\mathbf{r}$ on the target plane as a function of $\boldsymbol{\Delta}\mathbf{k}$ is achieved for
\begin{align}
\boldsymbol{\Delta}\mathbf{k}^{\rm opt}=\frac{3k\boldsymbol{\Delta}\mathbf{r}}{2L}.
\end{align}
This allows us to compare $\mathcal{C}^{\rm RT}$ with $\mathcal{C}^{\rm FP}$ not only when $\boldsymbol{\Delta}\mathbf{k}=\mathbf{0}$, but also in the generalized shift-tilt memory effect for the optimal isoplanatic patch.

It is worth emphasizing that the general expression of $\mathcal{C}^{\rm RT}(\alpha)$, given in Eq.~(\ref{correlation}), cannot be compared with $\mathcal{C}^{\rm FP}(\alpha)$ directly.
For $\tau<1$, one has to subtract the constant background contribution $I_0\exp(-\tau)$ from $\Gamma(\alpha)$ and then normalize it (see, e.g., the discussion in Ref.~\cite{Vellekoop_Optica4_2017}).
As shown in Eq.~(\ref{C-approx}), the constant background in the correlations depends on the shape of scatterers and is related to the ballistic light contribution and, hence, is beyond the FP model.

\section{Spatial correlations in thermally tunable magneto-optical media}

Let us study the shift-shift and tilt-tilt correlations in the framework of the rigorous Lorenz-Mie theory and the radiative transfer equation in the SAA.
Here, we are interested in a two-dimensional disordered medium composed of parallel cylinders normally irradiated with plane waves.
By a two-dimensional scattering medium we mean that the lengths of cylinders are much larger than both the incident wavelength and their diameters, so that most of the light is scattered on the plane perpendicular to the cylinder axis~\cite{bohren}.
In particular, the cylindrical geometry allows us to describe the application of an external magnetic field along the $z$ direction exactly, and it can be realized experimentally (see, e.g., Refs.~\cite{Kort-Kamp_PhysRevLett111_2013,Kort-Kamp_JOSAA31_2014} and references therein).
However, before studying a tunable metamaterial, it is worth presenting the dielectric case and comparing cylindrical and spherical scatterers, since the latter are widely investigated in spatial correlations~\cite{Raymer_PhysRevA62_2000,Judkewitz_NatPhys11_2015}.

\subsection{Optical memory effect for dielectric spheres and cylinders}
\label{sec3:dielectric-cylinders}

\begin{figure}[htpb!]
\includegraphics[width=\columnwidth]{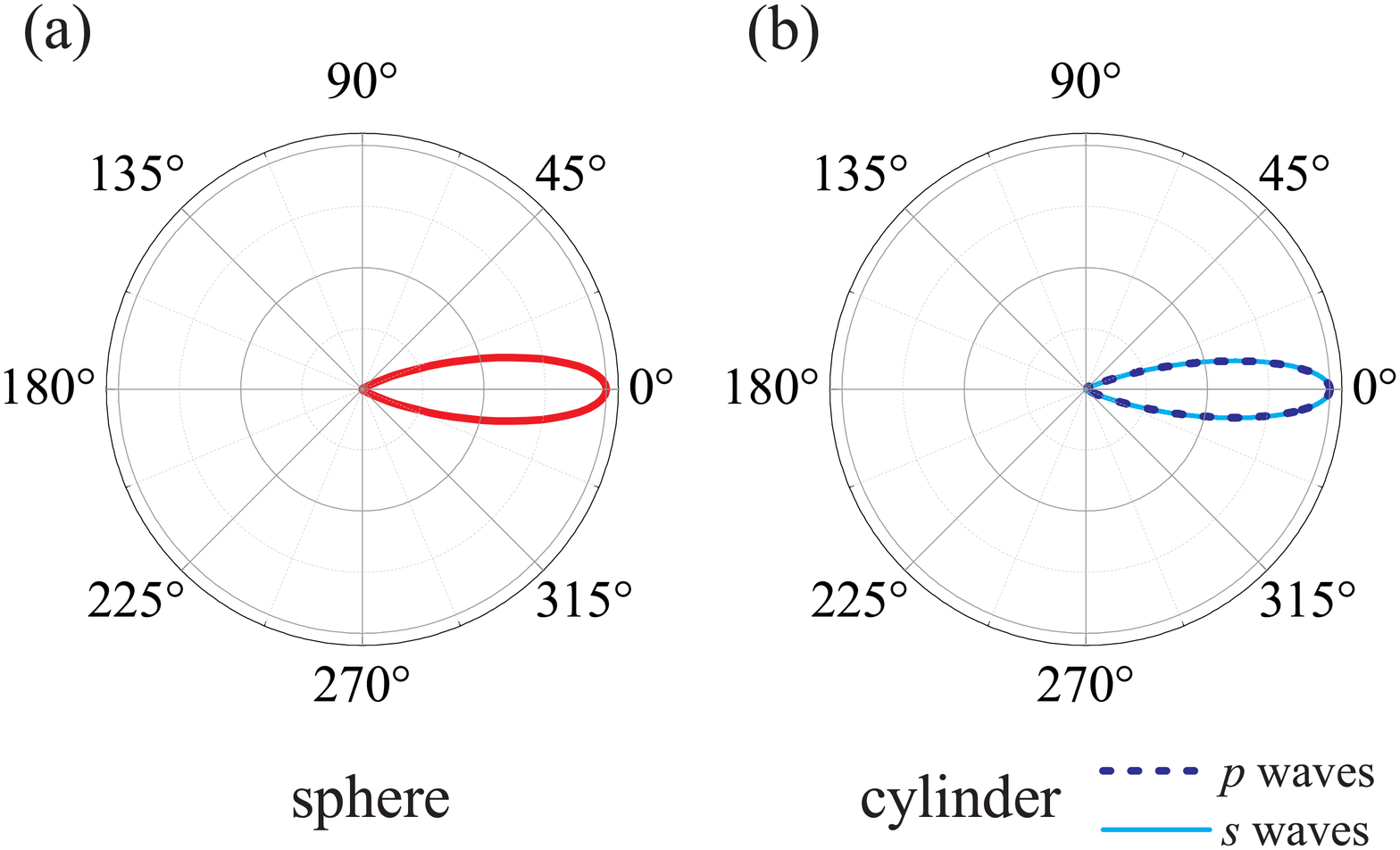}
\caption{
Phase function $p(\theta)$ calculated by the rigorous Lorenz-Mie theory for a silica ($\varepsilon_{\rm SiO_2}=2.25$) sphere (a) and a silica cylinder (b) embedded in agarose gel ($n_{\rm ag}=1.34$).
Both sphere and cylinder have radius $R=0.1$~mm and are normally irradiated with plane waves at a frequency $\omega=2\pi\times 2.4$~THz $(kR\approx 6.74)$.
The forward-peaked single scattering for this set of parameters allow us to use the small-angle approximation of the radiative transfer equation for small optical thicknesses~\cite{swanson}.
}\label{fig1}
\end{figure}

\begin{figure*}[htpb!]
\includegraphics[width=\columnwidth]{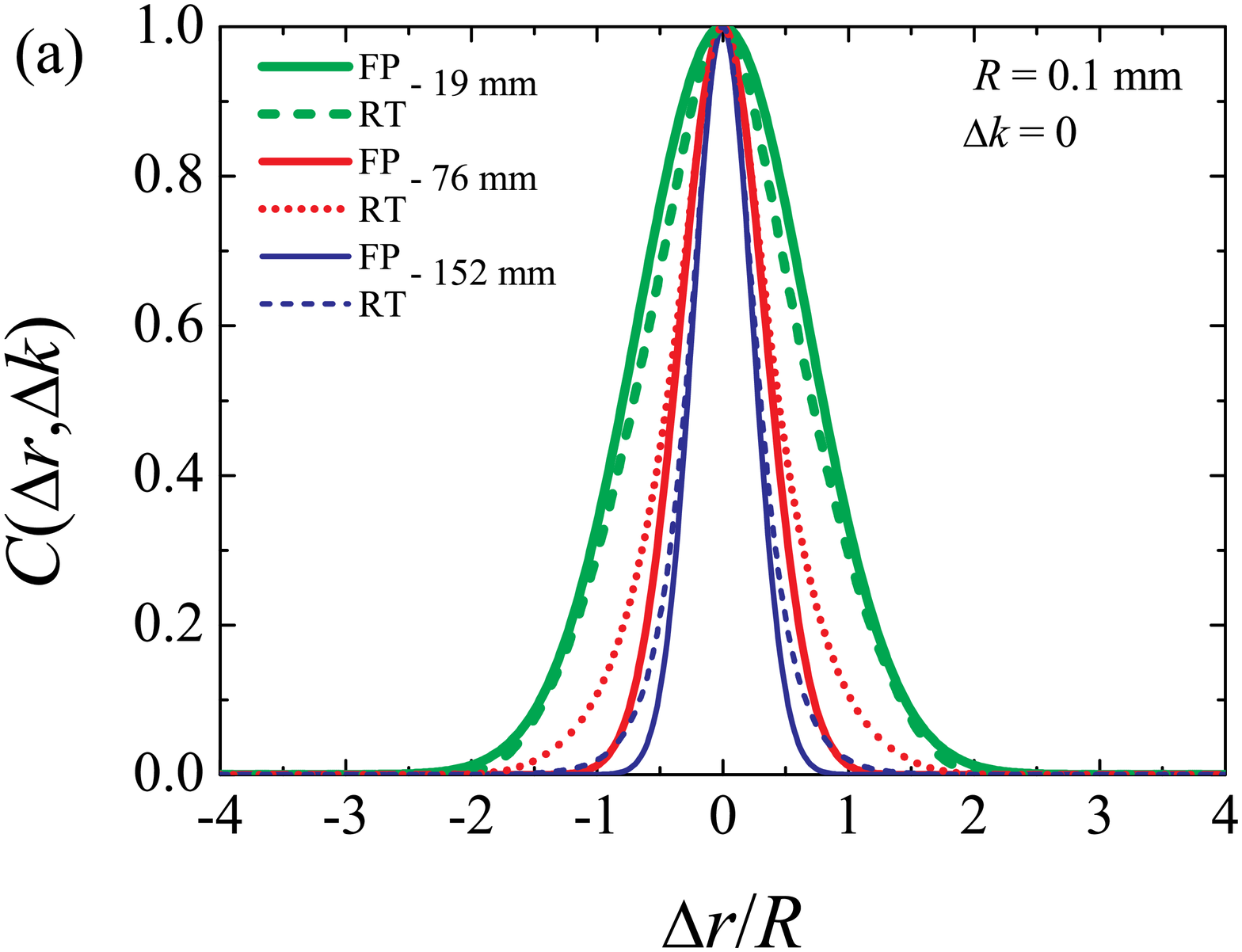}
\includegraphics[width=\columnwidth]{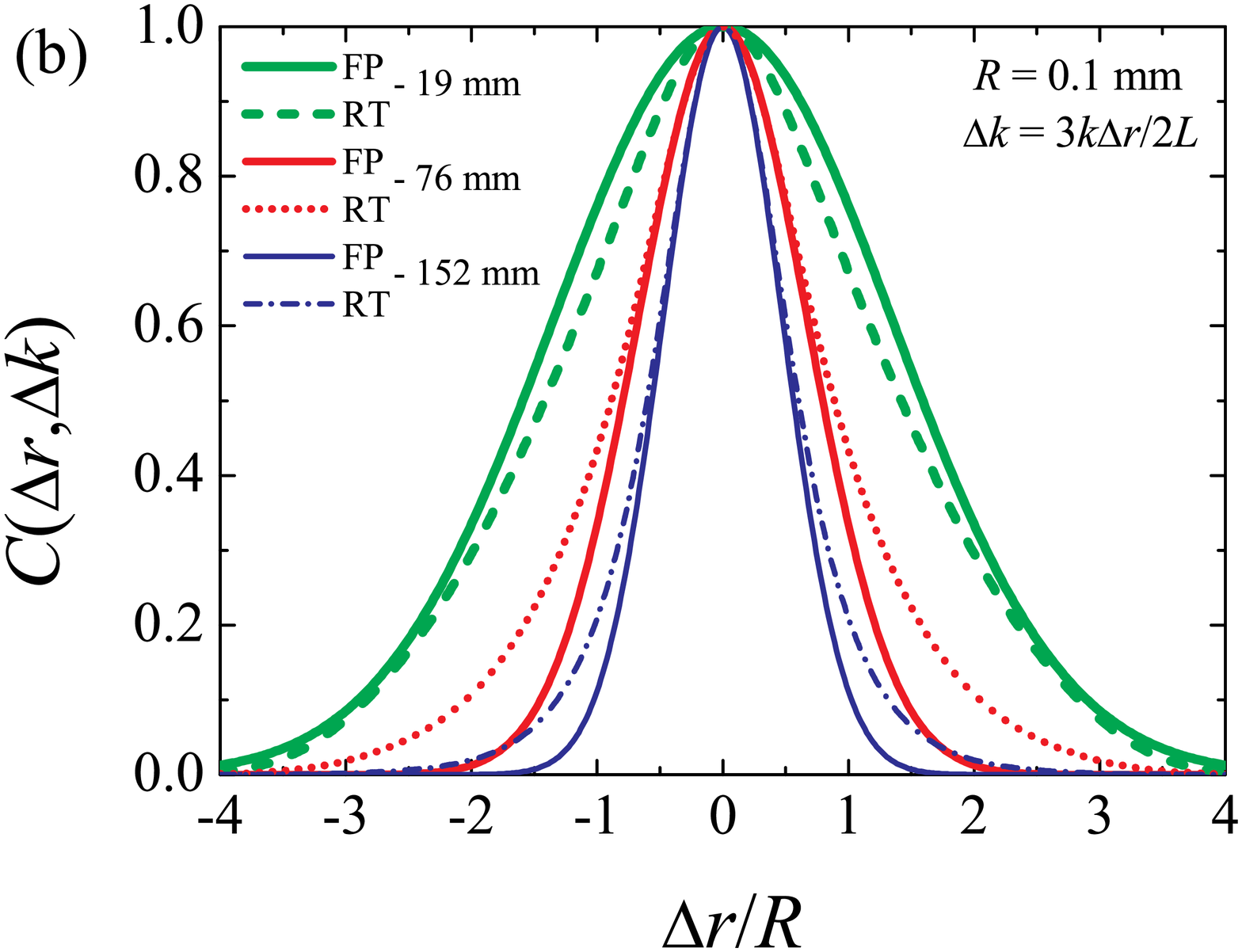}
\includegraphics[width=\columnwidth]{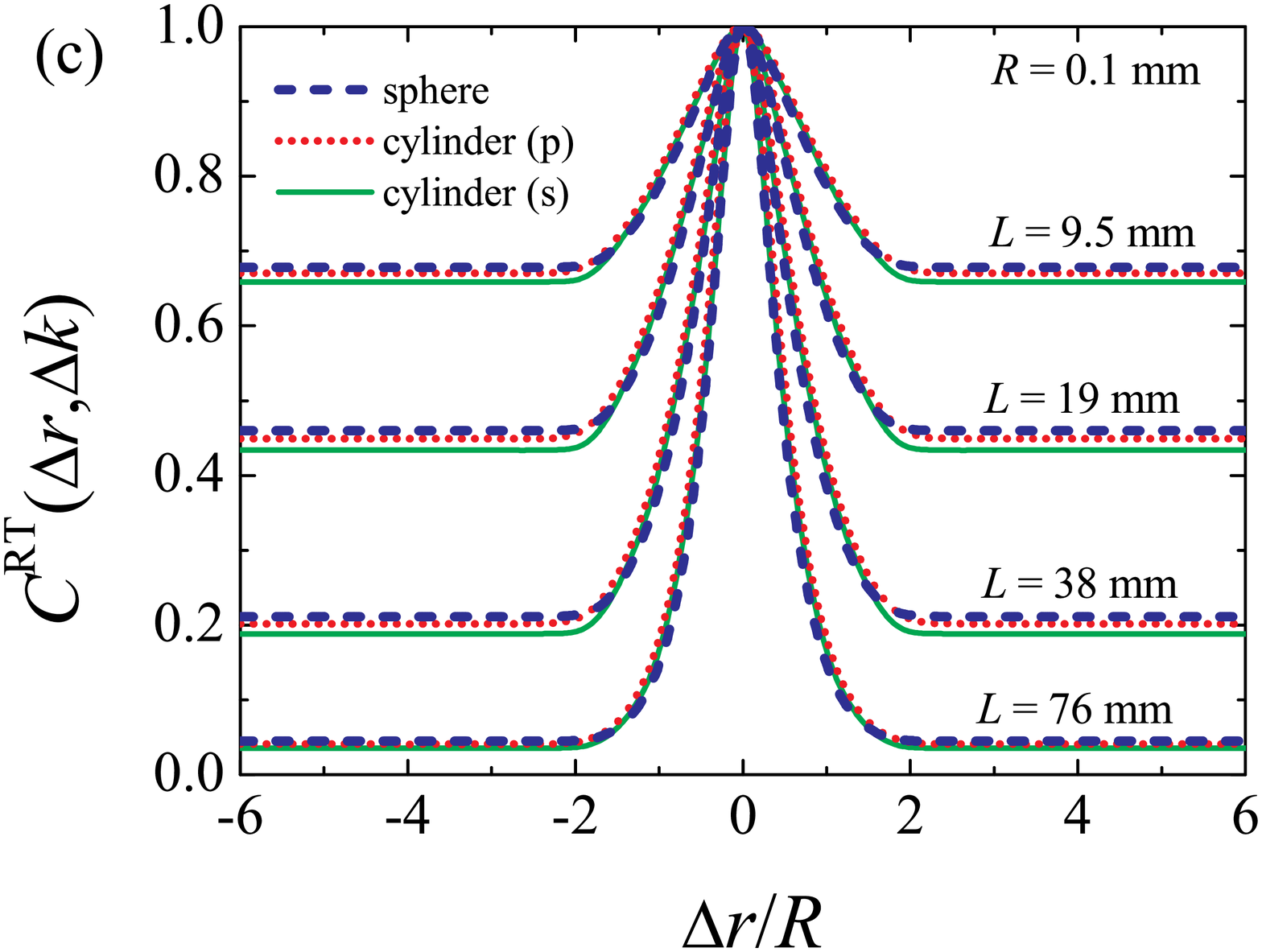}
\includegraphics[width=\columnwidth]{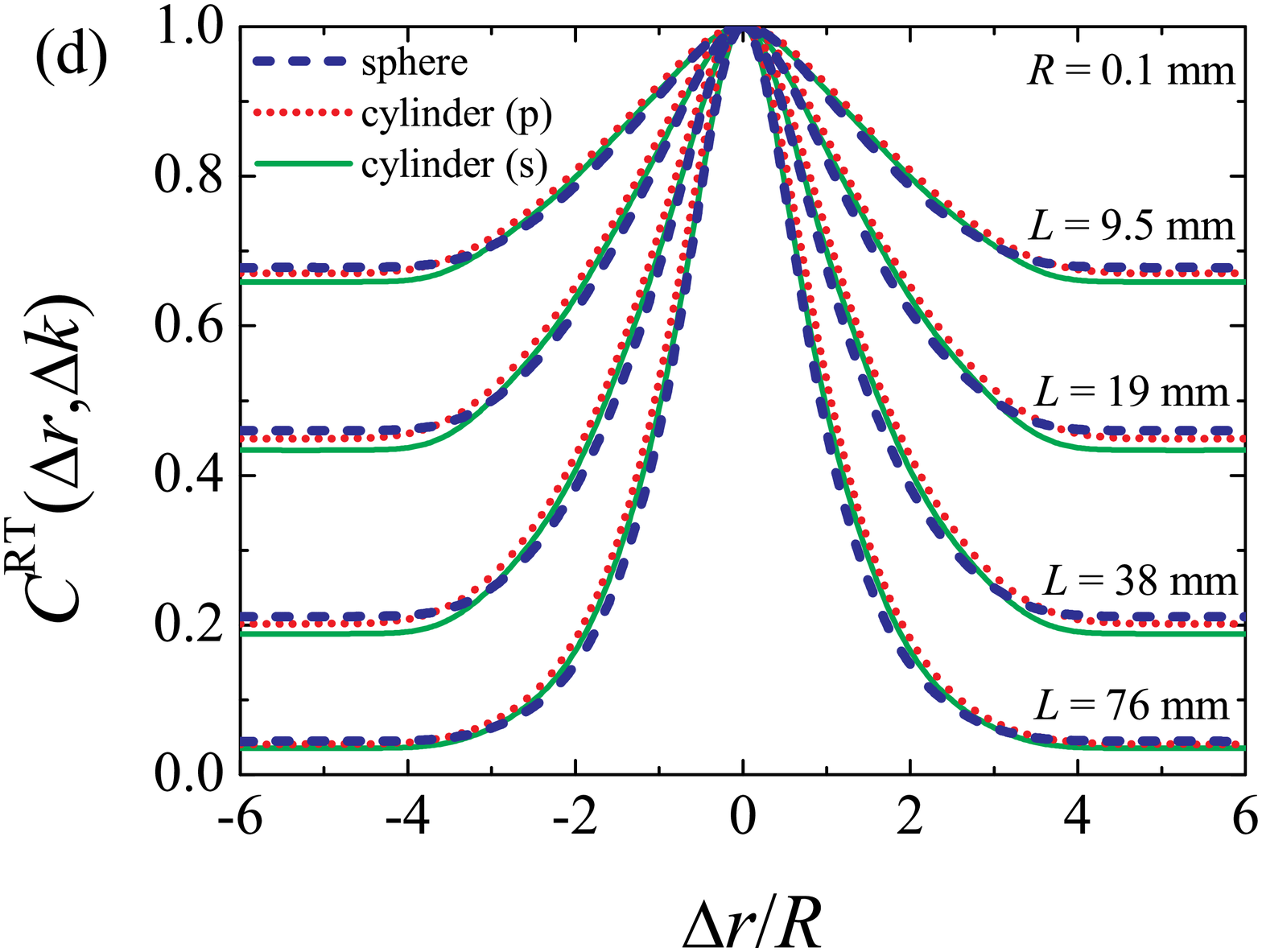}
\caption{
Spatial correlation function $\mathcal{C}(\boldsymbol{\Delta}\mathbf{r},\boldsymbol{\Delta}\mathbf{k})$ as predicated by the radiative transfer (RT) equation and by the Fokker-Planck (FP) equation for silica (SiO$_2$) monodispersed scatterers in agarose gel normally irradiated with \textit{p} or \textit{s} waves.
The phase functions are calculated by the rigorous Lorenz-Mie theory for SiO$_2$ spheres or cylinders with radius $R=0.1$~mm ($kR\approx6.74$) and packing fraction $f=0.45\%$, for various slab thicknesses $L$ and $\omega=2\pi\times 2.4$~THz.
(a) Comparison between $\mathcal{C}^{\rm RT}$ (with the background contribution subtracted) for spheres and $\mathcal{C}^{\rm FP}$ for $\Delta k=0$.
Solid lines correspond to $\mathcal{C}^{\rm RT}$; dashed lines, to $\mathcal{C}^{\rm FP}$.
(b) Shift-tilt memory effect for the optimal isoplanatic patch condition $\boldsymbol{\Delta}\mathbf{k} = 3k\boldsymbol{\Delta}\mathbf{r}/2L$.
For spheres or cylinders, (c) the shift-shift ($\Delta k=0$) and (d) the shift-tilt ($\boldsymbol{\Delta}\mathbf{k} = 3k\boldsymbol{\Delta}\mathbf{r}/2L$) correlations $\mathcal{C}^{\rm RT}$ are approximately the same.
}\label{fig2}
\end{figure*}

Within the framework of the Lorenz-Mie theory, the exact single-scattering phase function for spherical particles of radius $R$ interacting with plane waves is $p_{\rm s}(\theta) = (|S_1|^2+|S_2|^2)/[{2\pi (kR)^2}Q_{\rm sca}^{\rm s}]$, where $S_1$ and $S_2$ are dimensionless scattering amplitudes~\cite{bohren,hulst}.
Similarly, for cylinders of radius $R$ normally irradiated with $s$ waves ($\mathbf{E}||\hat{\mathbf{z}}$)~\cite{bohren,hulst}, one has  $p_{\rm c}(\theta) = |T_{1}|^2/[\pi kRQ_{\rm sca}^{\rm c}]$, where $T_1$ is the dimensionless scattering amplitude for $s$-polarized waves; for $p$ waves  ($\mathbf{E}\perp\hat{\mathbf{z}}$), one must consider $T_2$ instead of $T_1$ and the corresponding $Q_{\rm sca}^{\rm c}$~\cite{bohren,hulst}.

To compare spherical and cylindrical scattering geometries, we consider that the scatterers are made of silica (SiO$_2$) and have the same radius $R=0.1$~mm.
Here, we are interested in the terahertz (THz) frequency region, for which $\varepsilon_{\rm SiO_2}\approx 2.25$ is the silica permittivity.
The sphere (or cylinder) is embedded in a polymer binding matrix of refractive index $n_{\rm ag}=1.34$ (e.g., agarose gel)~\cite{Judkewitz_NatPhys11_2015}.
The scatterer is irradiated with plane waves at frequency $\omega=2\pi\times 2.4$~THz, which leads to $kR\approx 6.74$.

In Fig.~\ref{fig1}, we plot the normalized phase functions corresponding to a spherical [Fig.~\ref{fig1}(a)] and a cylindrical [Fig.~\ref{fig1}(b)] SiO$_2$ scatterer in agarose gel.
Clearly both $p_{\rm s}(\theta)$ and $p_{\rm c}(\theta)$ are forward-extended phase functions, with $\langle\cos\theta\rangle\approx0.937$ for the sphere, and $\langle\cos\theta\rangle\approx0.972$ ($p$ waves) and $\langle\cos\theta\rangle\approx 0.960$ ($s$ waves) for the cylinder.
For this set of parameters, we can apply the small-angle approximation of the RT equation.
According to Ref.~\cite{swanson}, the SAA is robust for moderate-sized parameters $(kR>6.6)$ if one considers small optical thicknesses $\tau$.

A system of monodispersed spheres (or parallel cylinders) of small optical thicknesses can be realized by considering a medium with a low density of scatterers ($f\ll1$).
Let us consider a packing fraction $f=0.45\%$ and large slab thickness compared to $R=0.1$~mm ($L\geq95 R$) in the THz frequency range.
For this set of parameters, the speckle is not yet fully developed implying a constant background in the spatial correlation function $\mathcal{C}(\boldsymbol{\Delta}\mathbf{r},\boldsymbol{\Delta}\mathbf{k})$.

To show the effect of the background in the correlations, we compare in Fig.~\ref{fig2} the Fokker-Planck  model with the solution of the radiative transfer  equation in the SAA.
In Fig.~\ref{fig2}(a), we plot the shift-shift memory effect ($\Delta k=0$), where $\mathcal{C}^{\rm RT}(k\Delta r)$ and $\mathcal{C}^{\rm FP}(k\Delta r)$ are calculated for a system of spherical scatterers.
The definition of the $\mathcal{C}^{\rm RT}(k\Delta r)$ was modified in order to subtract $I_0\exp(-\tau)$ from the transverse coherence function $\Gamma(\alpha)$~\cite{Vellekoop_Optica4_2017}.
As can be verified, the two theories show good agreement even for small optical thicknesses ($\tau<1$), as long as we neglect the ballistic (nonscattered) light contribution.
In particular, Fig.~\ref{fig2}(b) shows that the heuristic approach using the effective distance shift $\alpha$ into $\mathcal{C}^{\rm RT}(\alpha)$ agrees with the Fokker-Planck model.

Figures~\ref{fig2}(b) and \ref{fig2}(c) show the profiles of the spatial correlations $\mathcal{C}^{\rm RT}(\boldsymbol{\Delta}\mathbf{r},\boldsymbol{\Delta}\mathbf{k})$ including the ballistic light contribution as a function of the slab thickness $L$.
In particular, Fig~\ref{fig2}(b) can be compared with Fig.~3 in Ref.~\cite{Judkewitz_NatPhys11_2015}.
For both spherical and cylindrical scatterers, $\mathcal{C}^{\rm RT}(\alpha)$ in Fig.~\ref{fig2}(c) and Fig.~\ref{fig2}(d) decreases with increasing $L$ and approaches $\mathcal{C}^{\rm FP}(\alpha)$ plotted in Fig.~\ref{fig2}(a) and Fig.~\ref{fig2}(b), respectively.
In addition, as shown in Eq.~(\ref{relation}), note that the difference in the shift-shift correlation function for spheres versus cylinders is very small.
Indeed, for our set of parameters, we have verified that $\mathcal{C}_{\rm s}(\alpha)\approx \mathcal{C}_{\rm c}(\alpha)$.
This means that one can basically retrieve the spatial correlations of a three-dimensional anisotropically scattering medium from a two-dimensional system with equivalent parameters.

\subsection{Optical memory effect in magneto-optical disordered media}
\label{sec4:magneto-optical-cylinders}

The optical memory effect can be tuned by changing the optical thickness $\tau$ of an anisotropically scattering media.
As shown in Fig.~\ref{fig2}, $\mathcal{C}(k\Delta r)$ can be trivially enhanced by considering lower values of slab thickness $L$ within the SAA.
Indeed, for $\Delta k=0$, one can verify that~\cite{Judkewitz_NatPhys11_2015}
\begin{align}
\mathcal{C}(k\Delta r,\eta L)=[\mathcal{C}(k\Delta r,L)]^{\eta}.\label{C-pot}
\end{align}
This can be straightforwardly confirmed by the definition of $\mathcal{C}(\alpha)$; see Eqs.~(\ref{correlation}) and (\ref{C-approx}).
However, it is interesting to have external parameters that could change the memory effect without changing the geometry or density of particles.

\begin{figure}[htpb!]
\includegraphics[width=.9\columnwidth]{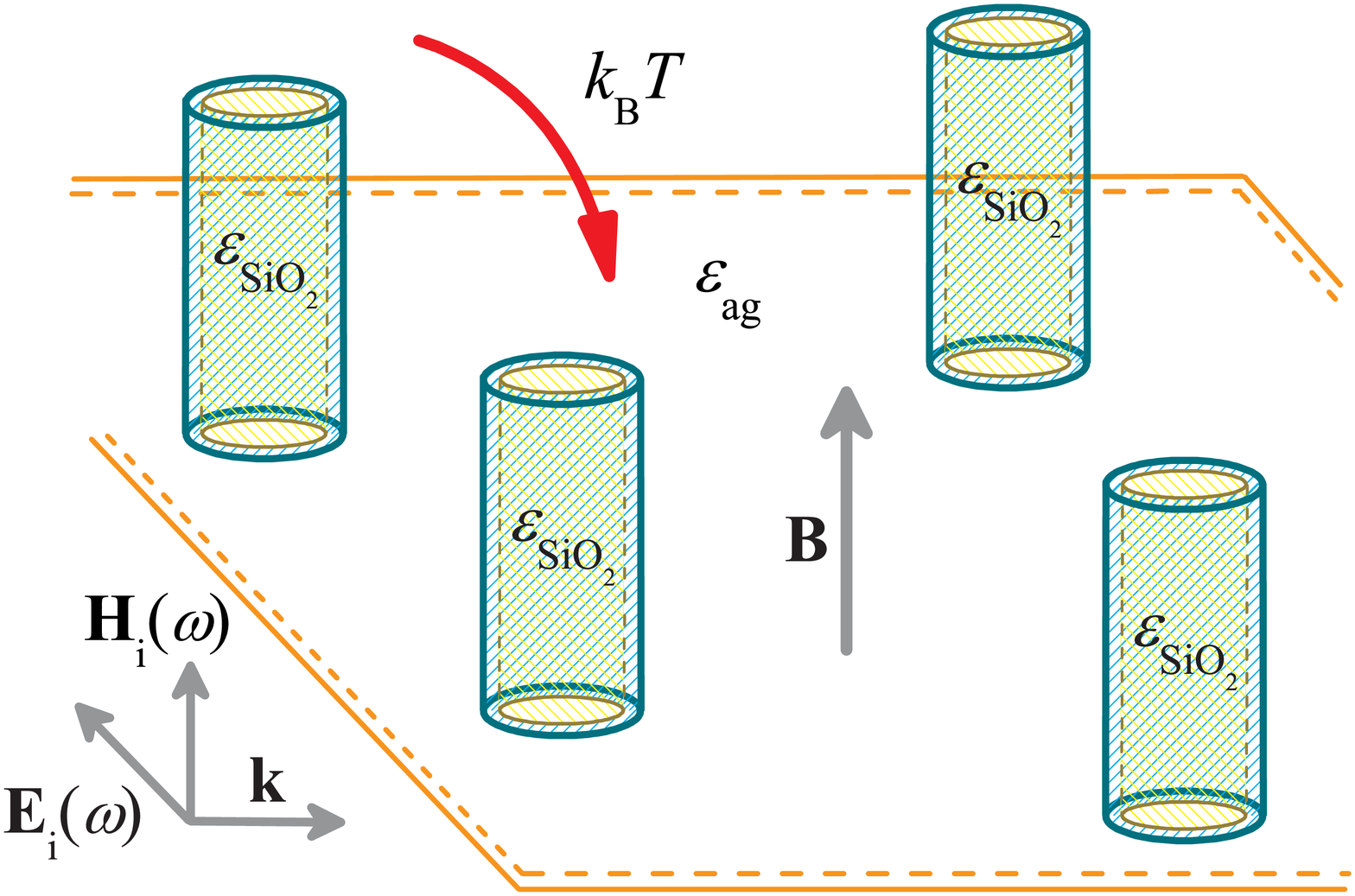}
\caption{Thick cylinders $(kR\gg1)$ irradiated with $p$ waves ($\mathbf{H}_{\rm i}||\hat{\mathbf{z}}$) at the presence of an external magnetic field $\mathbf{B}=B\hat{\mathbf{z}}$.
Scatterers are composed of a dielectric SiO$_2$ cylinder coated with a subwavelength magneto-optical shell of InSb embedded in agarose gel, with packing fraction $f\ll1$.
The InSb cylindrical shell is strongly dependent on the external magnetic field and the temperature $T$.
}\label{fig3}
\end{figure}

\begin{figure*}[htb!]
\includegraphics[width=\columnwidth]{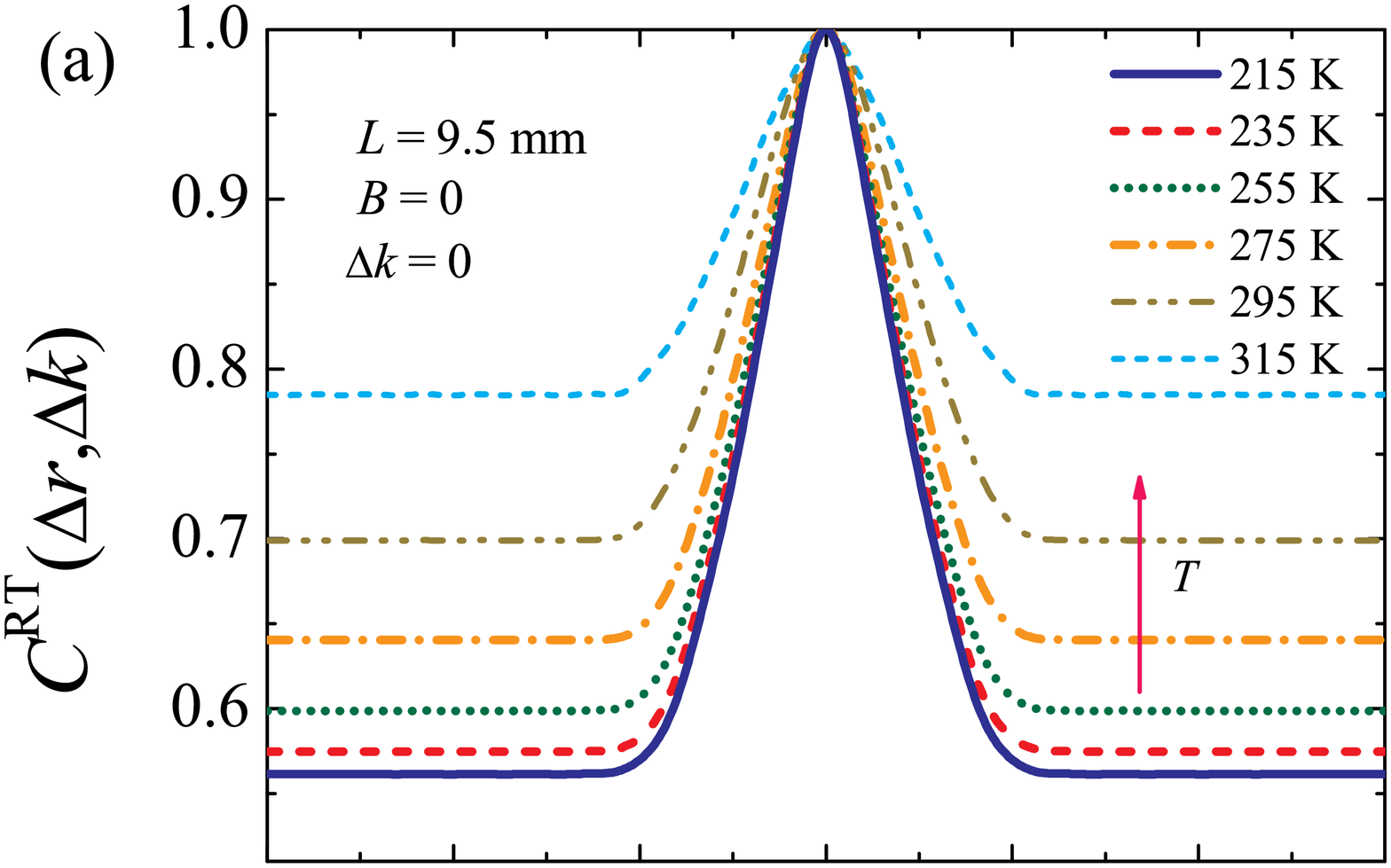}\vspace{-1.6cm}
\includegraphics[width=\columnwidth]{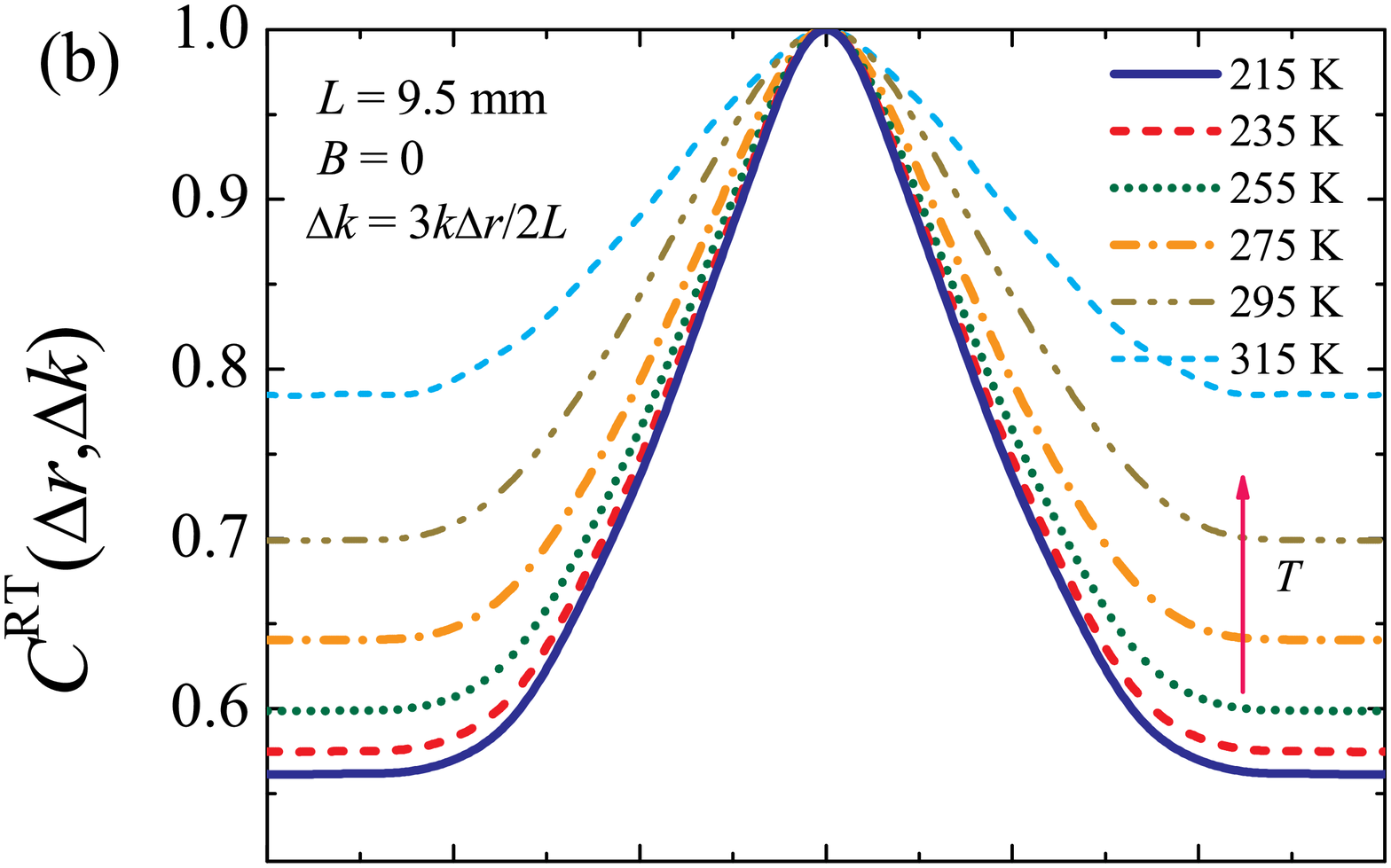}
\includegraphics[width=\columnwidth]{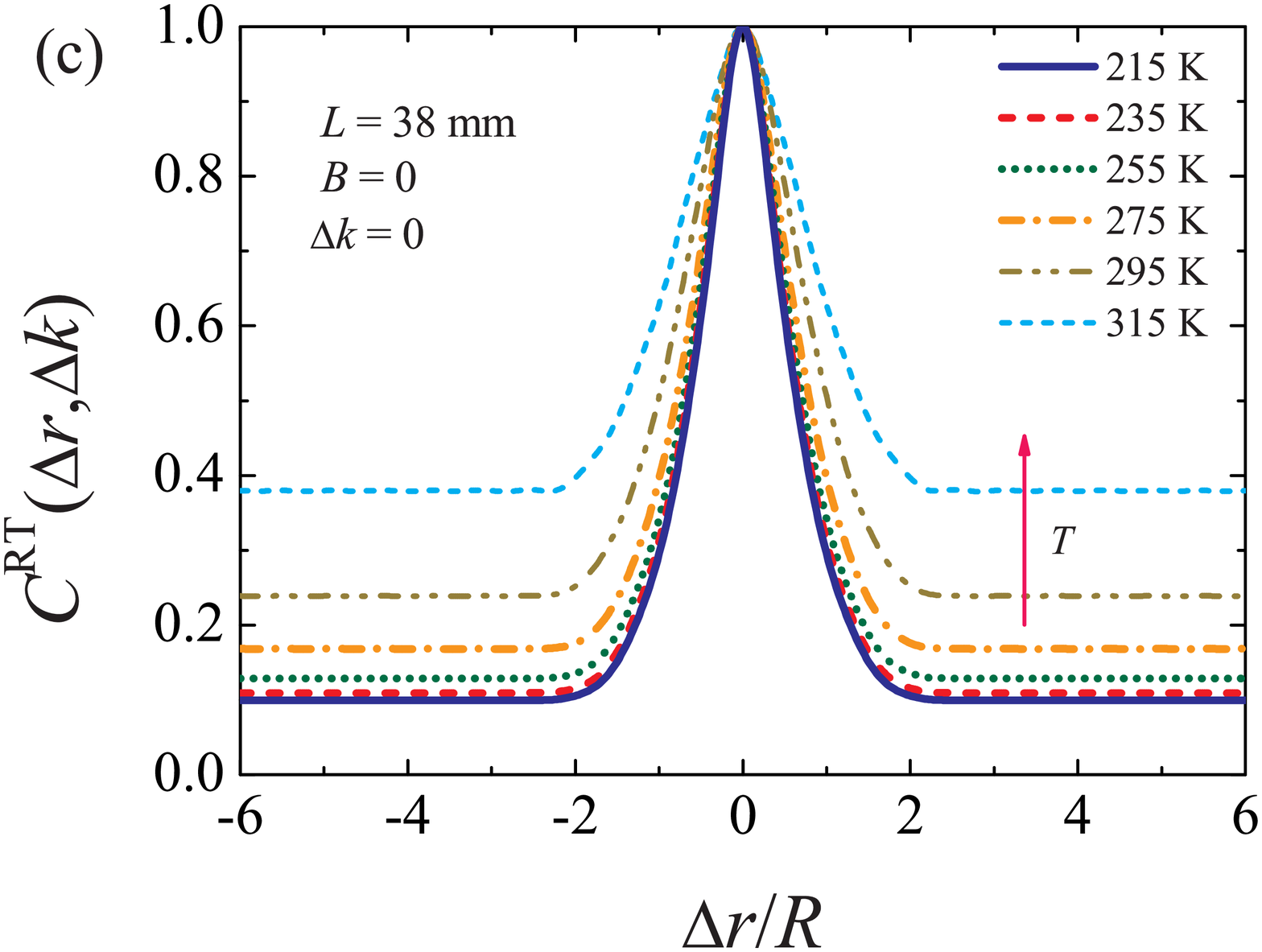}
\includegraphics[width=\columnwidth]{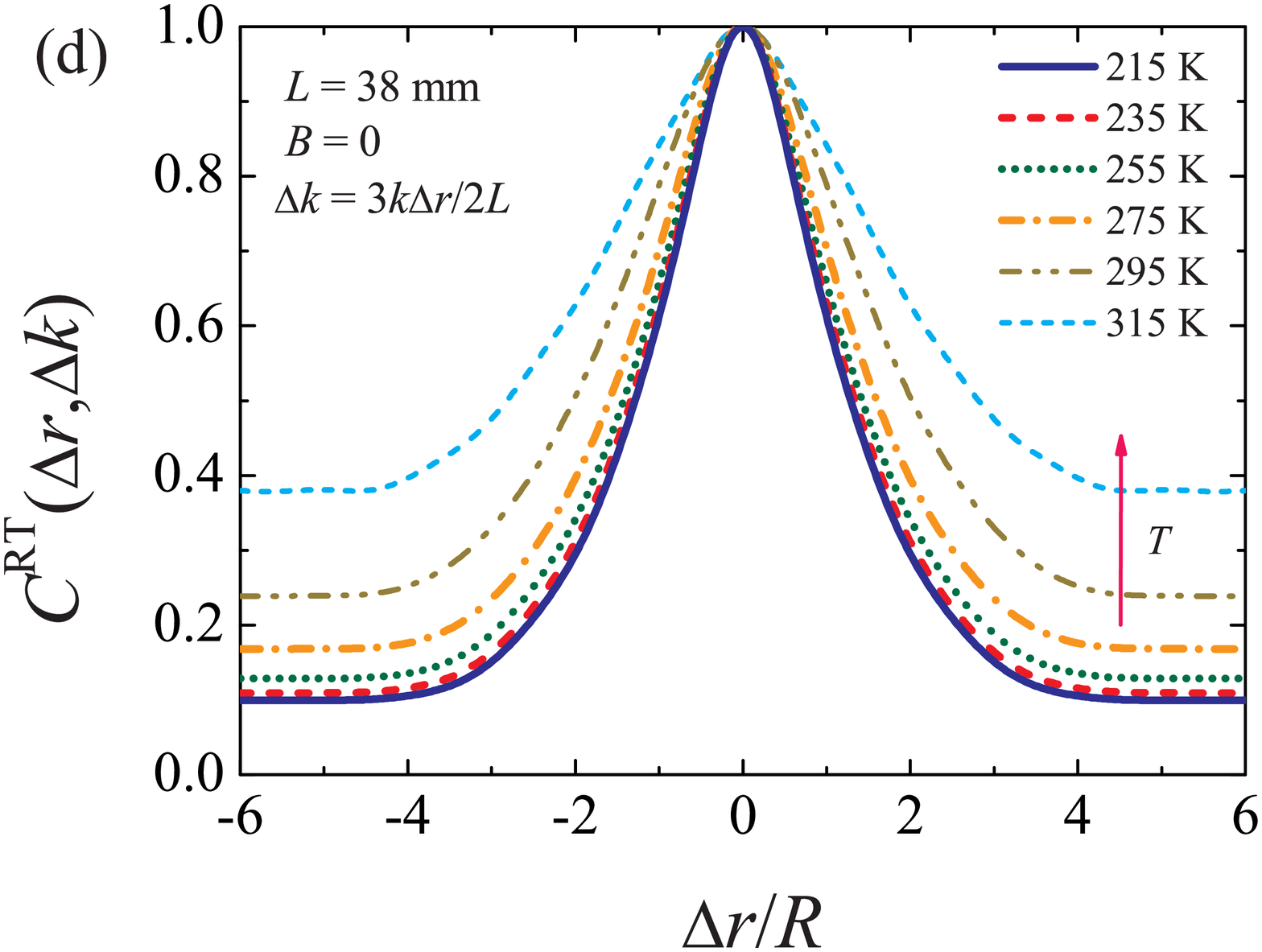}
\caption{
Spatial correlation function $\mathcal{C}(\boldsymbol{\Delta}\mathbf{r},\boldsymbol{\Delta}\mathbf{k})$ for monodispersed (SiO$_2$) core-shell (InSb) cylinders in agarose gel $(f=0.45\%)$ normally irradiated with $p$ waves ($\omega=2\pi\times 2.4$~THz).
The core-shell cylinder has radius $R=100~\mu$m $(kR\approx 6.74)$, where the SiO$_2$ core is coated with a InSb single layer of thickness $d=1~\mu$m $(kd\approx 0.0674)$.
The phase function is calculated by the rigorous Lorenz-Mie theory.
(a) Plot showing the increase in $\mathcal{C}$ as a function of the temperature $T$ for slab thickness $L=9.5$~mm and $\Delta k=0$, where the maximum memory effect within the validity of the SAA is found for $T=315$~K.
(b) Shift-tilt memory effect for the optimal isoplanatic patch condition $\boldsymbol{\Delta}\mathbf{k} = 3k\boldsymbol{\Delta}\mathbf{r}/2L$.
Even small correlations for a large slab thickness ($L=4\times9.5$~mm) can be considerably enhanced by increasing $T$ for (c) $\Delta k=0$ and (d) $\boldsymbol{\Delta}\mathbf{k} = 3k\boldsymbol{\Delta}\mathbf{r}/2L$.
}\label{fig4}
\end{figure*}

\begin{figure*}[htb!]
\includegraphics[width=\columnwidth]{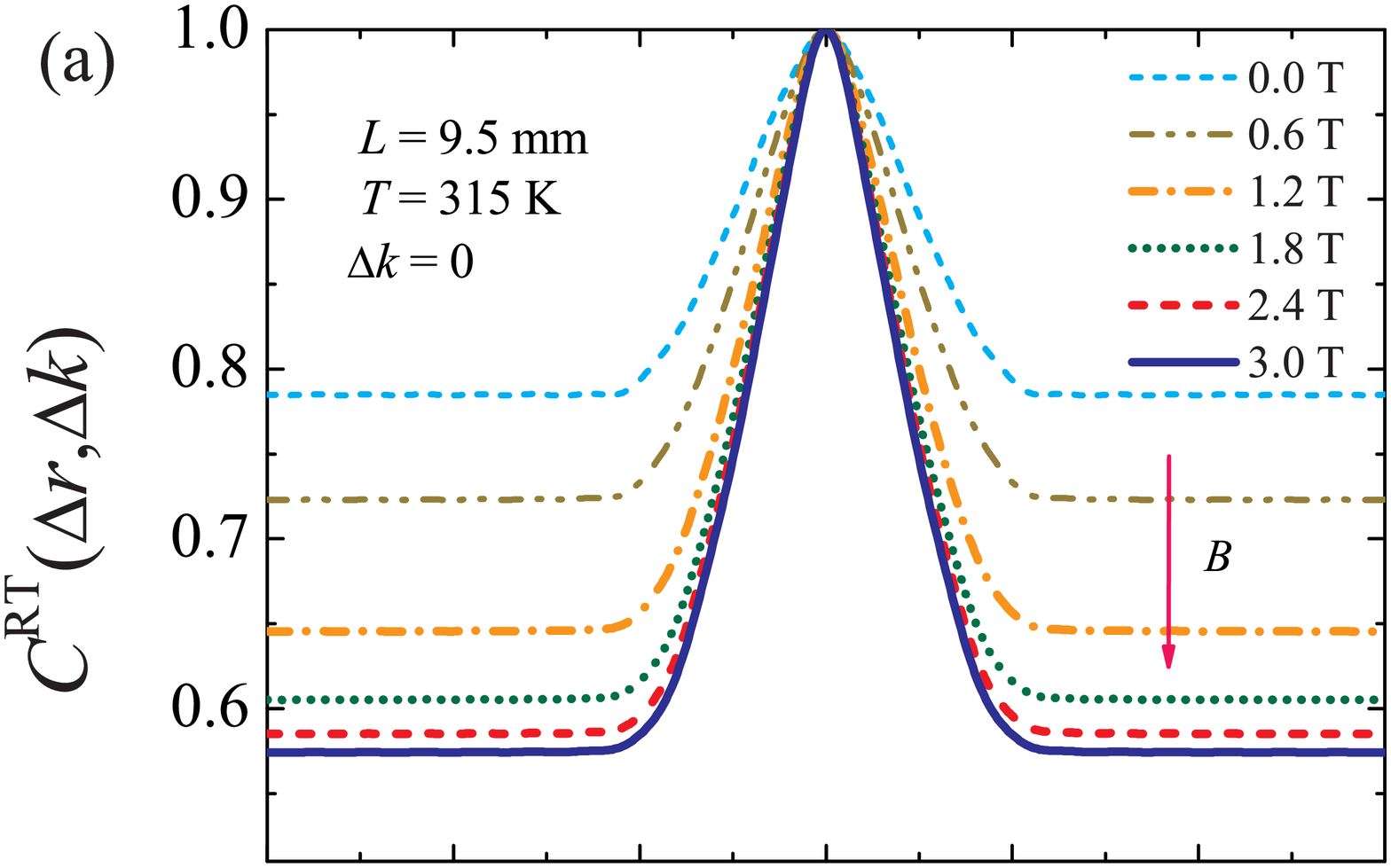}\vspace{-1.6cm}
\includegraphics[width=\columnwidth]{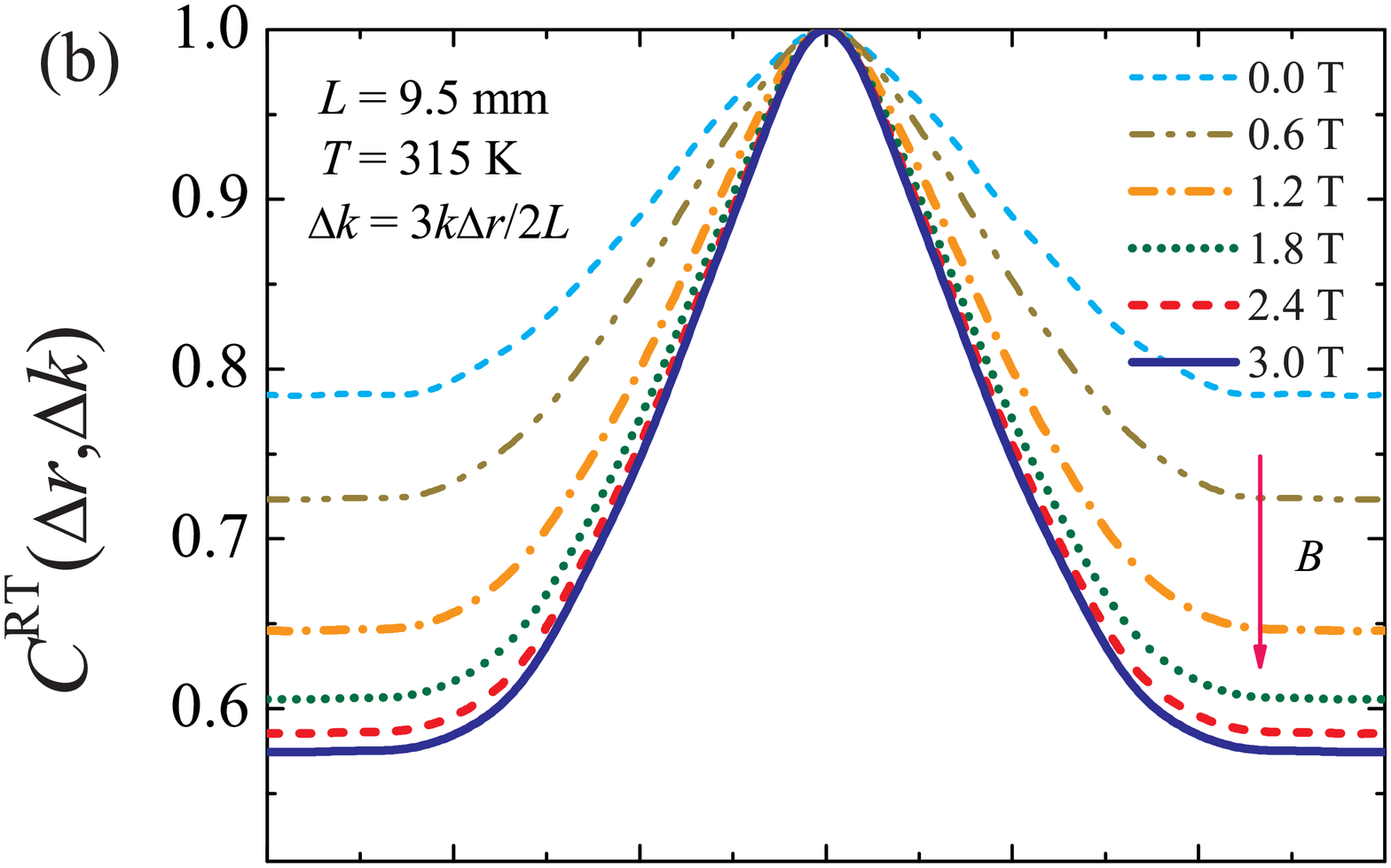}
\includegraphics[width=\columnwidth]{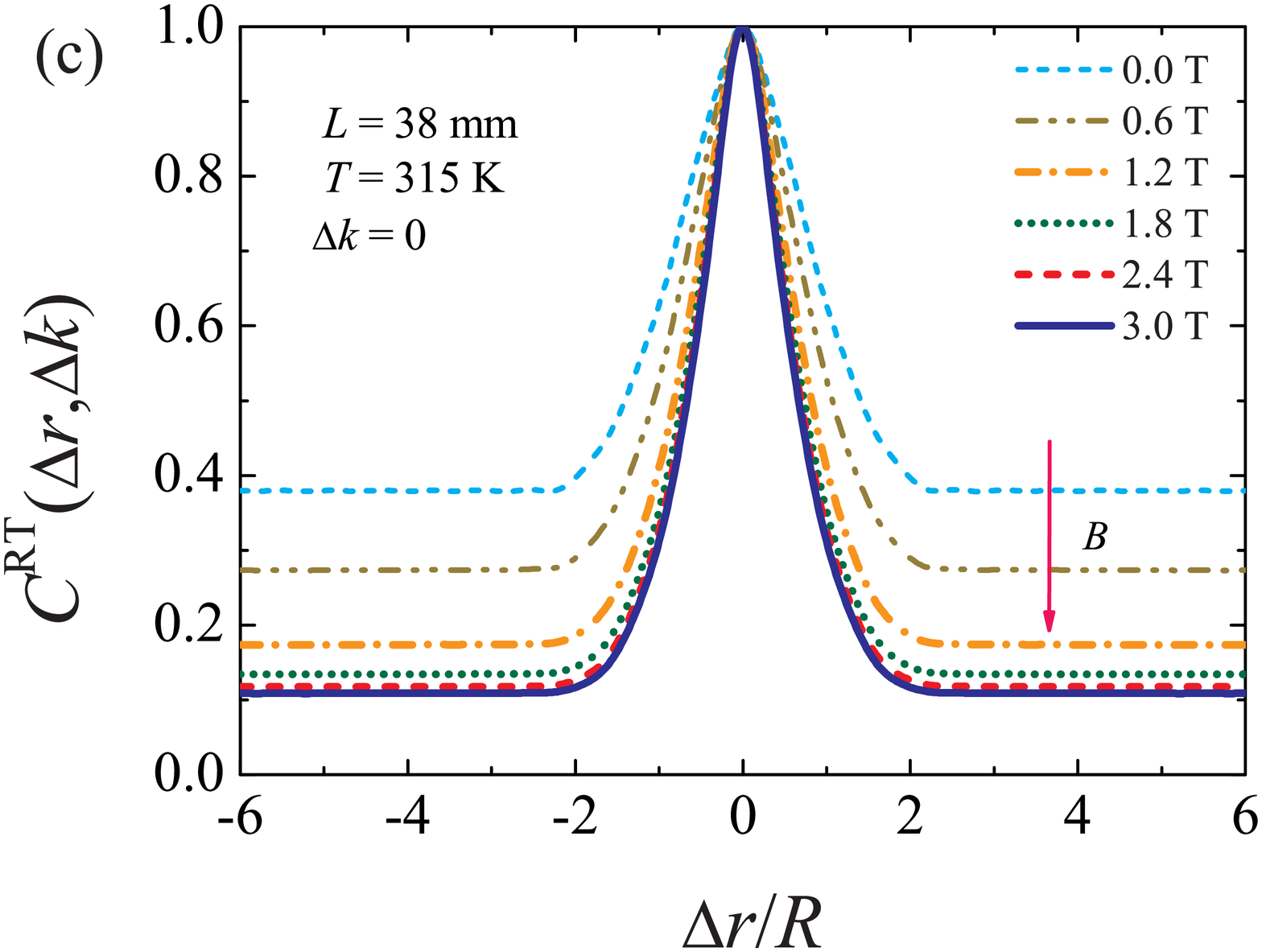}
\includegraphics[width=\columnwidth]{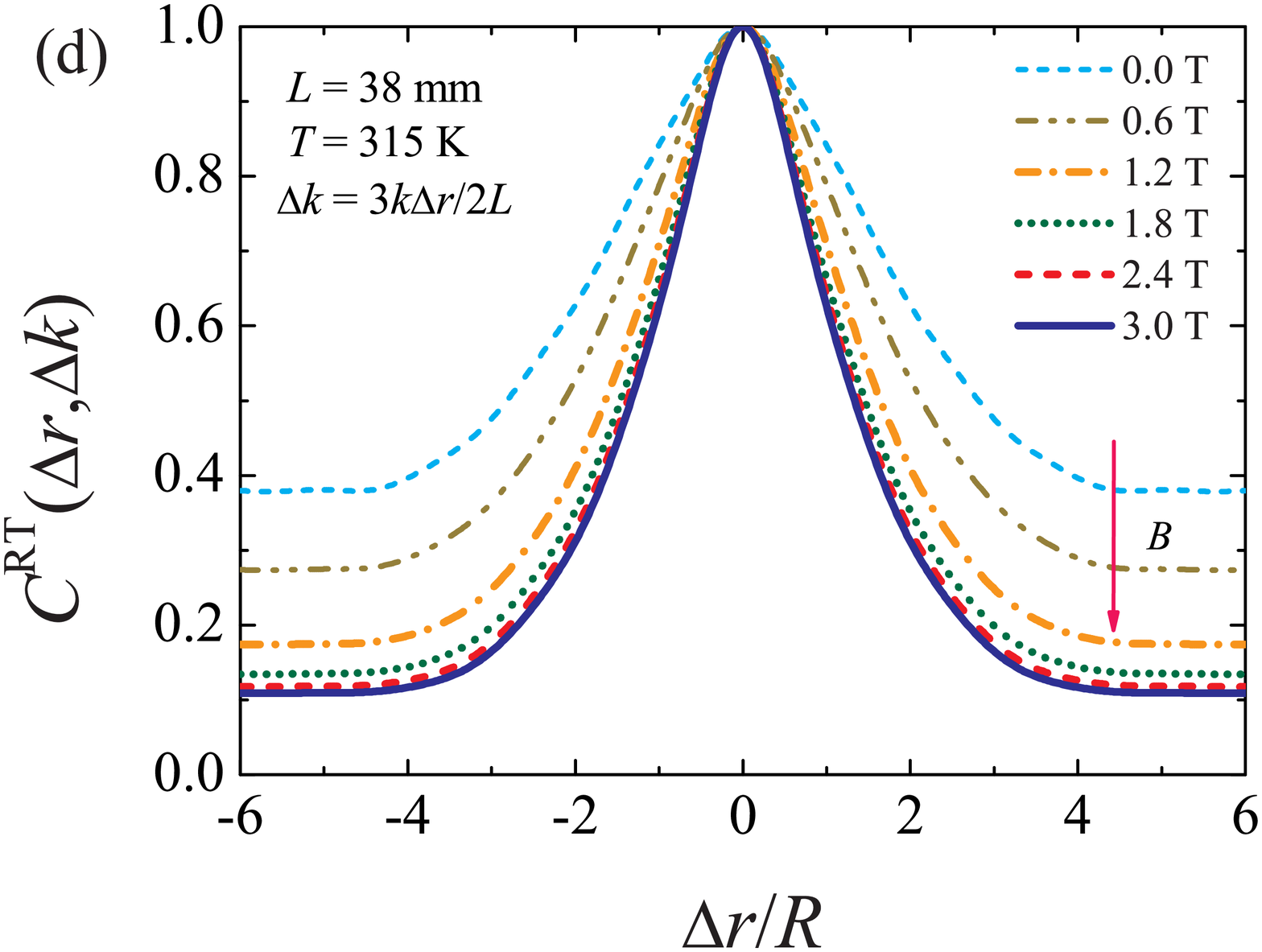}
\caption{
Shift-shift correlation function $\mathcal{C}(\boldsymbol{\Delta}\mathbf{r},\boldsymbol{\Delta}\mathbf{k})$ for monodispersed (SiO$_2$) core-shell (InSb) cylinders in agarose gel $(f=0.45\%)$ normally irradiated with $p$ waves ($\omega=2\pi\times 2.4$~THz).
The core-shell cylinder has radius $R=100~\mu$m $(kR\approx 6.74)$, where the SiO$_2$ core is coated with a InSb single layer of thickness $d=1~\mu$m $(kd\approx 0.0674)$.
The phase function is calculated by the rigorous Lorenz-Mie theory.
The plots show that the maximum correlation achieved for $T=315$~K in Fig.~(\ref{fig4}) can be decreased by the application of an external magnetic field $\mathbf{B}=B\hat{\mathbf{z}}$.
(a) $L=9.5$~mm and $\Delta k=0$.
(b) $L=9.5$~mm and $\Delta k=3k\Delta r/2L$.
(c) $L=4\times9.5$~mm and $\Delta k=0$.
(d) $L=4\times9.5$~mm and $\boldsymbol{\Delta}\mathbf{k} = 3k\boldsymbol{\Delta}\mathbf{r}/2L$.
}\label{fig5}
\end{figure*}

Here, we propose the use of thermally and magnetically tunable semiconductors with moderate permittivity in the THz frequency range~\cite{dai,chen,tiago-pra2016} to control the shift-shift and tilt-tilt memory effects.
The main idea is to use subwavelength coatings that are strongly dependent on the temperature and the external magnetic field, and exhibit a low refractive index.
These features could be achieved with semiconductor materials that are known to exhibit a high cyclotron frequency $\omega_{\rm c}$, such as InSb, InAs, HgTe, Hg$_{1-x}$Cd$_x$Te, PbTe, PbSe, PbS, and GaAs~\cite{zawadzki,raymond}.
For our calculations, we consider a subwavelength coating of indium antimonide (InSb)~\cite{madelung,cunninghan,zimpel}, whose dielectric tensor that fits experimental data can be found in Refs.~\cite{dai,chen,tiago-pra2016}.
In particular, we use the complete Lorenz-Mie theory for coated magneto-optical cylinders derived in Ref.~\cite{tiago-pra2016}.
The system geometry is illustrated in Fig.~\ref{fig3}.

Figure~\ref{fig4} shows the possibility of tuning the anisotropic memory effect by changing the temperature of the disordered system.
The system is composed of thick cylinders of radius $R=0.1$~mm irradiated by $p$ waves with $\omega=2\pi\times 2.4$~THz ($kR\approx 6.74$), as depicted in Fig.~\ref{fig3}.
Once again, as in Fig.~\ref{fig2}(c) and Fig.~\ref{fig2}(d), the cylinders are composed of SiO$_2$ in a polymer binding matrix of agarose gel.
However, we consider now that the cylinders are coated with a very thin, homogeneous subwavelength layer of InSb of thickness $d=1~\mu$m ($kd\approx 0.0674$).
Since the refractive index of InSb is moderate in the THz range for low temperatures and moderate magnetic fields, a subwavelength shell ($R/d=100$) guarantees that the effective refractive index of the whole scatterer is dominated by the refractive index of the SiO$_2$ core.
Indeed, despite the lossy InSb shell, we have verified that the single-scattering albedo is $\varpi_0^{\rm c}\approx1$.

In Fig.~\ref{fig4}(a) and Fig.~\ref{fig4}(b) we calculate $\mathcal{C}^{\rm RT}(\alpha)$ for the slab thickness $L=9.5$~mm.
Once again, we focus on the shift-shift memory effect $(\Delta k=0)$ and the shift-tilt memory effect under the optimal isoplanatic patch condition ($\boldsymbol{\Delta}\mathbf{k} = 3k\boldsymbol{\Delta}\mathbf{r}/2L$).
In the temperature range 215~K$<T<315$~K, we show unprecedented control of the spatial correlations in anisotropically scattering media $(0.916<\langle\cos\theta\rangle<0.974)$.
Without changing any geometrical parameter of the system as we have done in Fig.~\ref{fig2}, the spatial correlation is continuously enhanced as the temperature increases.
In particular, a greater variation in $\mathcal{C}^{\rm RT}(\alpha)$ is achieved in the range 275~K$<T<315$~K.
For $T>315$~K, the system exhibits $\langle\cos\theta\rangle<0.9$, and hence the SAA cannot be applied accurately.
For the range 150~K$<T<215$~K, we have verified that the spatial correlation remains approximately constant.
In particular, in Fig.~\ref{fig4}(c) and Fig.~\ref{fig4}(d) we consider $L=4\times9.5$~mm, which corresponds to a configuration of lower spatial correlation.
Even in this case, we can considerably enhance the spatial correlation by increasing the temperature.

Figure~\ref{fig5} shows that applying an external magnetic field $\mathbf{B}=B\hat{\mathbf{z}}$ has the opposite effect on $\mathcal{C}^{\rm RT}(\alpha)$ compared to increasing the temperature of the system.
Fixing $T=315$~K, i.e., the maximum spatial correlation obtained in Fig.~\ref{fig4} for our set of parameters, we can retrieve the spatial correlations of lower temperatures just by increasing $B$.
Indeed, the plots in Figs.~\ref{fig5}(a)--\ref{fig5}(d) for 0~T$<B<3$~T and $T=315$~K $(0.916<\langle\cos\theta\rangle<0.952)$ can be mapped in Figs.~\ref{fig4}(a)--\ref{fig4}(d), where 275~K$<T<315$~K and $B=0$.
For $B>3$~T, we have verified that the spatial correlation remains approximately constant.
Physically, this decreasing of the correlation as a function of $B$ can be explained by the breaking of degeneracy of the multipole scattering channels for $B\not=0$~\cite{tiago-pra2016}.

In Fig.~\ref{fig5}(c) and Fig.~\ref{fig5}(d) we repeat the same analysis as in Fig.~\ref{fig4}(c) and Fig.~\ref{fig4}(d) for $L=4\times 9.5$~mm, showing the effect of $\mathbf{B}\not=\mathbf{0}$ on a configuration of lower spatial correlation.
By comparing Fig.~\ref{fig4} and Fig.~\ref{fig5}, we show explicitly that the optical memory effect can be tuned by $T$ and $B$ to enhance or decrease the spatial correlation, respectively.
Indeed, for $B=0$, one can design configurations of low temperature $T_0$ and small slab thickness $L_0$ whose spatial correlations match a configuration of higher temperature $T>T_0$ and greater slab thickness $L>L_0$.
The same idea can be applied for $B\not=0$ and a fixed temperature, where a configuration in a low magnetic field $B_0$ with large slab thickness $L_0$ can reproduce the same shift-tilt memory effect of a configuration in a higher magnetic field $B>B_0$ with a smaller slab thickness $L<L_0$.

\section{Conclusion}
\label{sec5:conclusion}

 We have proposed a  material platform, based on core-shell metamaterials, to enhance and externally tune memory effects in disordered optical media.
In addition, we have provided further understanding of these effects by deriving an approximate expression for the generalized optical memory effect in the context of ballistic propagation of light.
Here, we have investigated  optical memory effects for cylindrical scatterers and have established a connection between the shift-shift memory effect for cylinders and spheres, which correspond to the case that has been treated so far.
In particular, for monodispersed large scatterers, the ballistic light contribution in the transverse coherence function is shown to depend only on geometric parameters of the system.
By considering an anisotropically two-dimensional scattering medium containing thick dielectric cylinders coated with subwavelength magneto-optical layers, we have demonstrated unprecedented control of the spatial correlations in disordered scattering media, and that cannot be achieved with naturally occurring optical media, such as biological tissues.
Indeed, we have shown that the shift-shift and tilt-tilt correlation in the THz regime for (SiO$_2$) core-shell (InSb) cylinders can be enhanced by increasing the temperature, and it can be decreased by applying an external magnetic field.
Altogether, the present findings demonstrate the versatility of coated metamaterial scatterers to control optical memory  effects, opening new vistas for the engineering of photonic disordered devices.

\section*{Acknowledgments}

The authors thank N. Papasimakis for useful discussions.
T.J.A. acknowledges Coordena\c{c}\~ao de Aperfei\c{c}oamento de Pessoal de N\'ivel Superior (CAPES; Grant No. 1564300) and S\~ao Paulo Research Foundation (FAPESP; Grant No. 2015/21194-3) for financial support.
A.S.M. held grants from Conselho Nacional de Desenvolvimento Cient\'{\i}fico e Tecnol\'ogico (CNPq; Grant No. 307948/2014-5).
F.A.P.  acknowledges The Royal Society-Newton Advanced Fellowship (Grant No. NA150208), CAPES, CNPq, and Rio de Janeiro Research Foundation (FAPERJ) for financial support.

\appendix*

\section{Single-scattering phase functions for large spheres and thick cylinders}
\label{appendix-phase}

Approximate phase functions for spheres and cylinders of radius $R$ in the limit of large scatterers compared to the wavelength ($kR\gg1$) are well-known~\cite{bohren,hulst}.
From the diffraction theory, large spheres provide phase functions of the form~\cite{hulst,ishimaru}
\begin{align}
p_{\rm s}(\theta) \approx \frac{1}{4\pi}\left(1+\cos\theta\right)^2\left[\frac{J_1(kR\sin\theta)}{\sin\theta}\right]^2.\label{sphere}
\end{align}
Over the angular region of interest, one usually takes $(1+\cos\theta)/2\approx1$ for $kR\sin\theta>10$ to a very good approximation~\cite{bohren}.
Here, we maintain this prefactor to highlight similarities between spherical and cylindrical scattering geometries.
Equation~(\ref{sphere}) leads to  $\widetilde{p}_{\rm s}(0) \approx 1 - J_0^2(\pi\beta/2) - J_1^2(\pi\beta/2)= 1-4/(\pi^2\beta)$ for $|\theta|\ll1$.
For an arbitrary $\alpha$, one has the well-known approximation~\cite{kokhanovsky2,kokhanovsky_book}
\begin{align}
\widetilde{p}_{\rm s}(\alpha)\approx\frac{2}{\pi}\left[\arccos\left(\frac{\alpha}{\beta}\right) - \frac{\alpha}{\beta}\sqrt{1-\left(\frac{\alpha}{\beta}\right)^2}\right]u(\alpha),\label{large-sphere}
\end{align}
where $\beta=2kR$, $u(\alpha\leq \beta)=1$ and $u(\alpha>\beta)=0$.
Note that Eq.~(\ref{large-sphere}) implies that the transverse coherence function $\Gamma(\alpha)$ depends only on $\tau_{\rm s}$ and $\varpi_0^{\rm s}$ for $\alpha>\beta$.

We can obtain a similar result regarding cylindrical scatterers.
The phase function of infinitely long cylinders with $kR\gg1$, upon normal incidence of radiation with arbitrary polarization, can be approximated by the diffraction theory~\cite{bohren},
\begin{align}
p_{\rm c}(\theta) \approx \frac{1}{4\pi}(1+\cos\theta)^2\left[\frac{\sin(kR\sin\theta)}{\sqrt{kR}\sin\theta}\right]^2,\label{cylinder}
\end{align}
which leads to $\widetilde{p}_{\rm c}(0)\approx (2/\pi)\int_0^{\pi kR}{\rm d}\xi(\sin\xi/\xi)^2\to1$ as $kR\to\infty$.
Also, one has
\begin{align}
\widetilde{p}_{\rm c}(\alpha) \approx \frac{2}{\pi}\int_0^{ \pi\beta/2}{\rm d}\xi\frac{\sin^2\xi}{\xi^2}\cos\left(\frac{2\alpha}{\beta}\xi\right),
\end{align}
where $\widetilde{p}_{\rm c}(\alpha>\beta)\approx 0$.
Indeed, we have verified that Eq.~(\ref{large-sphere}) is also a good approximation for $\widetilde{p}_{\rm c}(\alpha)$.
By inspection of Eqs.~(\ref{trans-sphere}) and (\ref{sphere}), this approximate equality between spherical and cylindrical cases follows from the asymptotic expansions $J_0(z)\approx\sqrt{2/\pi z}\cos(z-\pi/4)$ and $J_1(z)\approx\sqrt{2/\pi z}\sin(z-\pi/4)$ for $|z|\gg1$~\cite{bohren}.

\end{document}